SHORT COMMUNICATION

# Evaluation of non-ionic and zwitterionic detergents as membrane protein solubilizers in two-dimensional electrophoresis

Sylvie Luche[1], Véronique Santoni[2] and Thierry Rabilloud[1]*

1: CEA- Laboratoire de Bioénergétique Cellulaire et Pathologique, EA 2943, DRDC/BECP, CEA-Grenoble, 17 rue des martyrs F-38054 GRENOBLE CEDEX 9 France

2: UMR5004 Biochimie et physiologie moléculaire des plantes, place Viala, 34060 Montpellier cedex 1, France

Running title: detergents for 2D electrophoresis of membrane proteins

*: To whom correspondence should be addressed

Correspondence :

Thierry Rabilloud, DRDC/BECP, CEA-Grenoble, 17 rue des martyrs, F-38054 GRENOBLE CEDEX 9
Tel (33)-4-38-78-32-12
Fax (33)-4-38-78-51-87
e-mail: Thierry.Rabilloud@cea.fr

Abbreviations:

ASB 14: tetradecanoylamido propyl dimethyl ammonio propane sulfonate. Brij 30: tetraethylene glycol mono dodecyl ether. Brij 56: decaethylene glycol mono hexadecyl ether. Brij 58: docosaethylene glycol mono hexadecyl ether. Brij 78: docosaethylene glycol mono octadecyl ether. Brij 96: decaethylene glycol mono oleyl ether. C7 BzO: 3-(4-heptyl) phenyl 3-hydroxy propyl dimethyl ammonio propane sulfonate. C13E10: decaethylene glycol mono tridecyl ether. Cymal-6: 6-cyclohexyl-hexyl-_-maltoside. RBC: red blood cell. Tween 40: polyoxyethylene sorbitan monopalmitate

SUMMARY

The solubilizing power of various nonionic and zwitterionic detergents as membrane protein solubilizers for 2D electrophoresis was investigated. Human red blood cell ghosts and Arabidopsis thaliana leaf membrane proteins were used as model systems. Efficient detergents could be found in each class, i.e. with oligooxyethylene, sugar or sulfobetaine polar heads. Among the commercially-available nonionic detergents, dodecyl maltoside and Brij56 proved most efficient. They complement the more classical sulfobetaine detergents to widen the scope of useful detergents for the solubilization of membrane proteins in proteomics.

Analysis of membrane proteins remains a major challenge for proteomics techniques based on 2D electrophoresis. For this reason, alternative methods based either on the use of SDS electrophoresis [1], or on the separation of digestion peptides and not of proteins [2, 3] have been described. Although these methods have proven successful for the identification of membrane proteins, they do not achieve the combination of quantitative analysis and separation of protein variants available through the use of 2D electrophoresis. There is therefore still a wide interest for the solubilization of membrane proteins under the conditions prevailing in isoelectric focusing. For quite a long time, membrane proteins could not be solubilized adequately under these conditions [4]. However, the introduction of thiourea as a chaotrope in addition to urea [5] and the introduction of dedicated zwitterionic detergents [6, 7] has allowed to display some membrane proteins on 2D gels. Nevertheless, it is quite clear that these chemicals do not represent a universal solution to the problem of membrane protein solubilization [8]. While SDS is known to solubilize almost any protein, the situation is very different under the low-salt, electrically-neutral detergent conditions prevailing under IEF conditions. In this case, the electrostatic repulsive effect brought by the ionic polar heads of SDS is not available, resulting in much poorer solubilization. Thus, it can be foreseen that various membrane proteins will not solubilize adequately with a single detergent under non-ionic conditions. In addition, several of the dedicated detergents published so far (e.g. in [7] ) are not commercially available, which precludes their general use by the scientific community. This situation prompted us to evaluate several nonionic, commercially-available detergents for their efficiency in membrane protein solubilization. For the sake of easy comparison with the previously-described zwitterionic detergents, systems already investigated were used, i.e. human red blood cell membranes [7], and *A. thaliana* leaf membranes [6], which were prepared according to those publications. The leaf membranes were incubated in the presence of Triton X-100 (ratio 1:1) with a final concentration 0.2 % (w/v), during 10 min at 4 °C, under gentle shaking. After centrifugation at 120000 g during 15 min the pellet was resuspended in the storage buffer and centrifugated at 120000 g during 15 min. This ensured partial delipidation and extraction of many extrinsic membrane proteins. The final membrane pellets (red blood cell or leaf) were solubilized in the desired extraction medium (urea 7M, thiourea 2M, DTT 20 mM, carrier ampholytes 0.4-1%%, and detergent 2-4%). The zwitterionic detergents ASB14 and C7BzO were synthetised as described previously [6, 7]. The glycolipid detergents dodecyl and tetradecyl melibionamide were prepared according to published methods [9]. The other detergents were commercially available and purchased from standard laboratory suppliers . Glucopon and Plantacare were available from Fluka. Alkyl maltosides (dodecyl or tetradecyl) were at least 98% pure. Pure anomers only (α or β ) were used.

A constant amount of protein, as determined on the membrane preparations, was solubilized for each condition tested. After ultracentrifugation (250000g, 30 minutes) to remove insoluble material, the supernatant was used for 2D electrophoresis. Consequently, the 2D patterns reflect both the ability of the chaotrope-detergent mixture to extract efficiently the proteins of interest and its ability to prevent losses during the 2D electrophoresis process.

IEF was performed with home-made immobilized pH gradients (linear gradient from 4 to 8) [10], or with commercial 3-10 gradients. The IPG strips were rehydrated to a 3%T final acrylamide concentration overnight in 7M urea and 2M thiourea containing 20mM DTT and 0.4% carrier ampholytes 3-10 (Pharmalytes) and the detergent of interest (2% final concentration except for CHAPS: 4%). The protein sample was mixed with the rehydration solution for in-gel application of the sample [10]. The following running conditions were then used: from 0 to 300 V in 1 min, 300 V for 3 hours, from 300 V to 3500 V in 1 hour, 3500 V for 20 hours.

After the IEF run, the IPG gel strips were incubated at room temperature for 10 min in 6 M urea, 30% (w/v) glycerol, 2% (w/v) SDS, 13 mM DTT, 0.125 M Tris, 0.1 M HCl . The second equilibration step was carried out for 5 min in the same solution, excepted that DTT was replaced by iodoacetamide 2.5% [11]. The second dimension (SDS-PAGE) was carried out on homogenous running gels (10%T) without a stacking gel. The gels were silver-stained according to Sinha *et al*. [12]. The chloride-carbonate exchanger (also named band III) was identified by mass spectrometry [7], while the H+ATPase and aquaporin from *A. Thaliana* were identified by immunoblotting.

Because of its abundance in the red blood cell membrane, which are also quite easy to prepare, band III was chosen as a first model to evaluate the solubilization obtained with various detergents in 2D gel electrophoresis. Typical results are shown on the figures. Because band III is a protein with 12 putative transmembrane domains, it is not detected in classical 2D maps using CHAPS as a detergent, as shown on figure 1A. However, dedicated zwitterionic detergents achieved proper solubilization and analysis of band III on 2D gels (Figure 1B). The fuzzy appearance of the band III spot is typical. It is probably linked to the heavy modifications reported on this protein (glycosylation, palmitylation) which already give a fuzzy band in SDS electrophoresis [7]. Figure 1 also shows the results obtained with detergents bearing a sugar-derived polar head, and a linear alkyl tail. The Tween family proved completely unsuccessful, and no band III was seen with Tween 40 (Fig. 1C). Tween 20 and 80 gave the same result (no band III solubilization). Oppositely, the maltoside detergents gave very encouraging results, and both _- and _-dodecyl maltosides performed well (Fig. 1D and E). Tetradecyl maltoside and Cymal-6 also gave the same results (not shown). Due to the high prices of these detergents, cheaper substitutes were also tested, such as Glucopon and Plantacare, which are mixtures of alkyl glycosides and maltosides. These detergents proved completely unsuccessful (fig 1F), as did the dodecyl and tetradecyl melibionamides [9].

Results obtained with detergents bearing an oligoethylene glycol polar head are shown on figure 2. The detergents tested varied in both the length of the polar head and of the linear alkyl tail. Triton X-100, which bears an oligoethylene glycol polar head but a complex tert-octylphenyl hydrophobic part, was also included in the tests. Here again, the results were quite variable from one detergent to another. While Brij 30 (Fig. 2A) proved unable to solubilize band III, Brij 58 (Fig2C) and Brij 78 (Fig. 2D) gave little solubilisation and poor focusing. Brij 96 (Fig. 2E) proved able to solubilize to limited amount of band III, which was however correctly focused. Brij 56 (Fig. 2B), Triton X-100 (Fig. 2F) and C13E10 (not shown) were shown to be as efficient as zwitterionic detergents or glycosidic detergents.

In order to alleviate any bias which could be linked to the nature of band III, another membrane preparation (A. thaliana leaf) was used as a control. It had been previously shown that this preparation contains high amounts of aquaporins and H+ATPase, which could be displayed on 2D gels if correct detergents were used [6]. Figure 3 shows typical results obtained with various classes of detergents on this membrane preparation. While ASB14 is clearly superior to C7BzO for band III [7], C7BzO is clearly more efficient for this plant membrane preparation (Fig. 3C vs 3B). As to the nonionic detergents, Brij 78 was clearly inefficient( Fig 3E), while Brij 56 was intermediate between ASB14 and C7BzO(Fig 3D vs 3B and 3C). Dodecyl maltoside was however clearly the most efficient detergent of the panel, as judged by the amounts of aquaporins and H+ ATPase solubilized and focused (Fig. 3F).

The solubilization efficiency depends not only on the membrane protein, but also on the lipid content and on the treatments of the membrane preparation prior to final solubilization [13]. In this context, it must be noted that the plant membrane preparation is partially delipidated with Triton X-100, while the red blood cell membrane preparation is not delipidated. This may explain, at least in part, the differences in detergent efficiency observed on the two preparations.

The situation is therefore quite complicated and it is advisable to be able to use several detergents to choose the one(s) performing best with the proteins or biological system of interest. On the one hand, several detergents which proved efficient for membrane protein solubilization are either commercially unavailable (e.g. C7BzO) or available on a limited basis and at a rather high price (e.g. ASB14). On the other hand, there are many nonionic detergents, which are commercially available at various prices.

Sugar-based detergents provided quite efficient solubilization in both systems. Their efficiency seems however somewhat dependent on batch to batch variation (not shown) and they are quite expensive chemicals. The purity of these chemicals seems of paramount importance, as shown by the inefficiency of mixtures (e.g. Plantacare, Fig. 1F) and by the batch to batch variations.

Oligoethylene glycol-based detergents are efficient even though they are almost only mixtures, but their efficiency is variable. It does not seem to vary with batches but seems to depend on their hydrophilic-lipophilic balance (HLB). Detergents with a either a low HLB (e.g. Brij 30, HLB = 9.7) or a high one (e.g. Brij 58, Brij 78 and Tween 40, HLB 15.7, 15.3 and 15.6 resp.) are inefficient. Oppositely, detergents with a medium HLB (Brij 56, Triton X100 or C13E10, HLB 12.9, 13.5 and 14.1 resp.) are efficient. This does not seem to hold true for detergents with an oleyl lipophilic part (Brij 96, HLB= 12.4). Such an optimum in the HLB also holds true for membrane solubilization under native conditions [14]. Some discrepancies in detergent efficiency seem however to be brought by the chaotrope used for denaturation. Under native conditions, Brij 30 and Brij 96 show a high efficiency [14], which is not the case with the urea-thiourea mixture (this work). This also seems to hold true for the Triton X100-CHAPS comparison. It has long been described that CHAPS is far superior to Triton X100 when urea alone is used as chaotrope [15]. However, none of these detergents achieved solubilization of membrane proteins [4]. When the urea-thiourea chaotropic mixture is used, Triton X100 becomes much more efficient than CHAPS for solubilizing membrane proteins.

As a conclusion, the optimal choice of a detergent still remains largely empirical. However, we describe here several alternatives to dedicated sulfobetaine detergents. Dodecyl maltoside seems the most efficient alternative, but it is even more expensive than sulfobetaines. For cheaper alternatives, Triton X-100 or Brij 56 are good choices. However, we sometimes experienced detergent precipitation with Brij 56 in the concentrated urea-thiourea mixture, and C13E10 appears to be a versatile choice.

ACKNOWLEDGEMENTS:

TR wants to thank the CNRS for personal support

LEGENDS TO FIGURES

Figure 1: 2D electrophoretic separation of red blood cell ghosts proteins
150 μg of red blood cell ghosts proteins were loaded on the 2D gels. The first dimension is a 4-8 linear pH gradient, and the second dimension a 10% acrylamide gel. Band III is indicated by an arrow. The proteins are extracted and focused in a solution containing 7M urea, 2M thiourea, 20 mM DTT, 0.4% carrier ampholytes and A: 4% CHAPS; B: 2% ASB14; C: 2% Tween 40; D: 2% _-dodecyl maltoside; E: 2% _-dodecyl maltoside, F: 2% Plantacare.

Figure 2. 2D electrophoretic separation of red blood cell ghosts proteins
Same conditions as in figure 1, except that the detergents used for extraction and focusing are:
A: 2% Brij 30; B: 2% Brij 56; C: 2% Brij 58; D: 2% Brij 78; E: 2% Brij 96; F: 1% Triton X100

Figure 3. 2D electrophoretic separation of A. thaliana leaf plasma membrane proteins
60μg of leaf plasma membrane proteins were loaded on the 2D gels. . The first dimension is a 3-10 linear pH gradient, and the second dimension a 10% acrylamide gel. H+ ATPase (AHA), aquaporin monomer (PIP1) and aquaporin dimer (PIP2) are indicated by arrows. The proteins are extracted and focused in a solution containing 7M urea, 2M thiourea, 20 mM DTT, 0.4% carrier ampholytes and A: 4% CHAPS; B: 2% ASB14; C: 2% C7BzO; D: 2% Brij 56; E: 2% Brij 78; F: 2% _-dodecyl maltoside

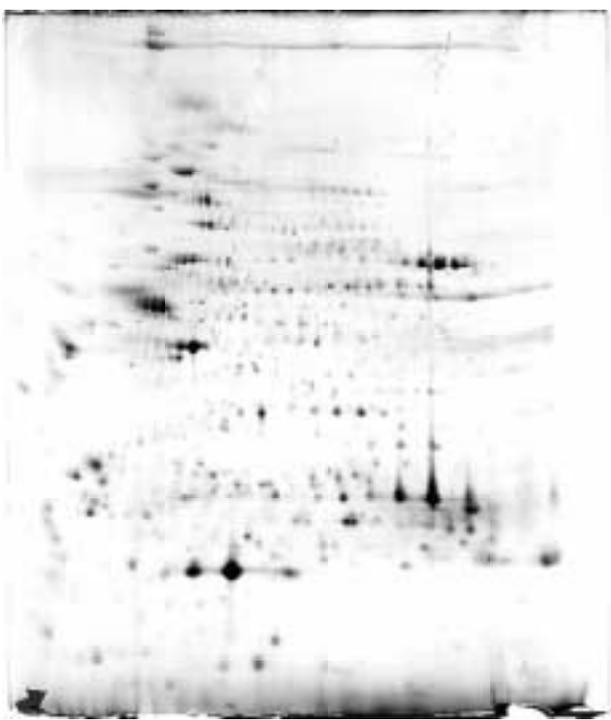
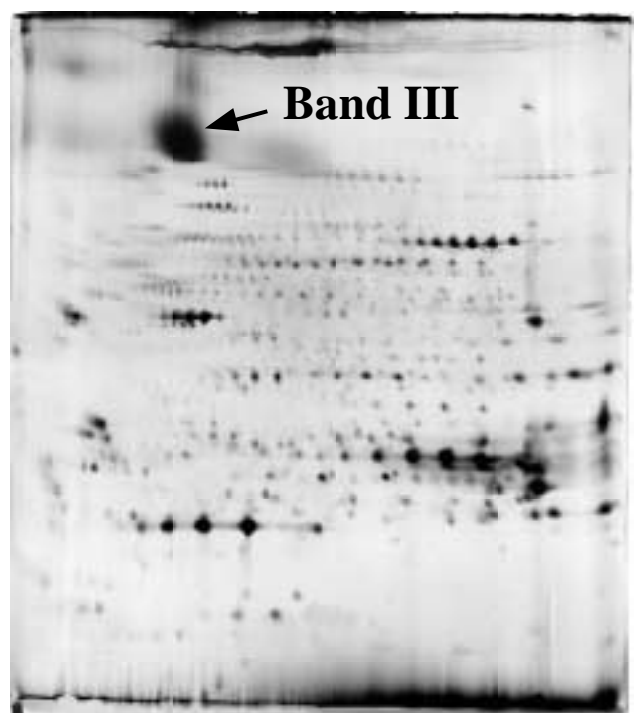
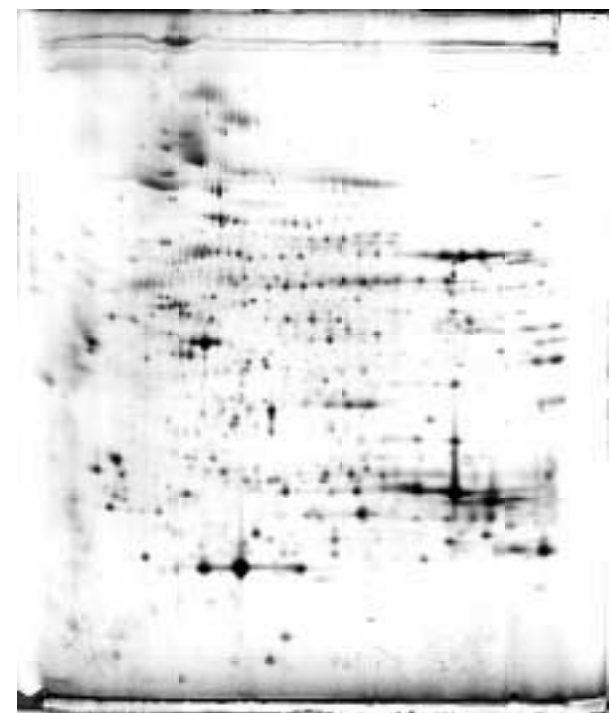
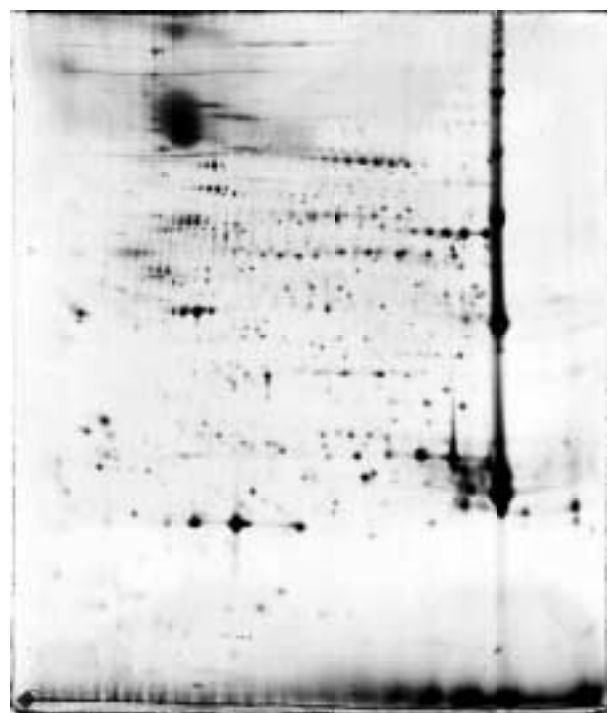
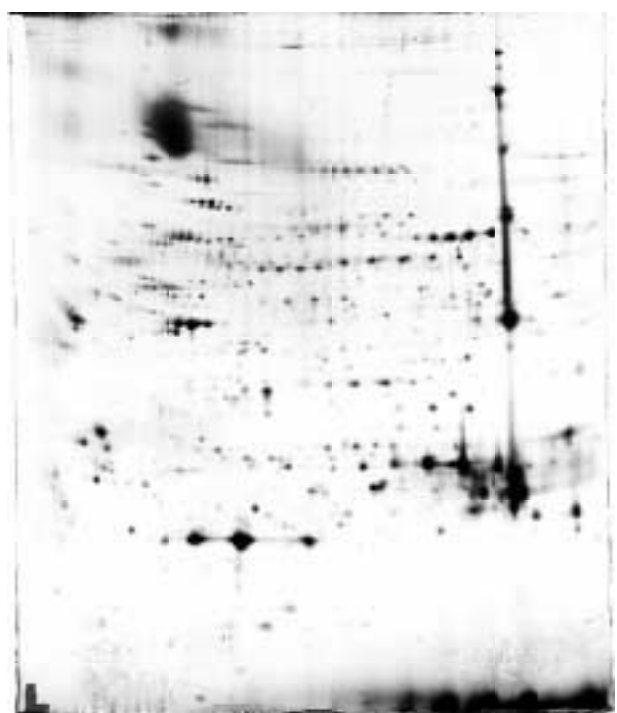
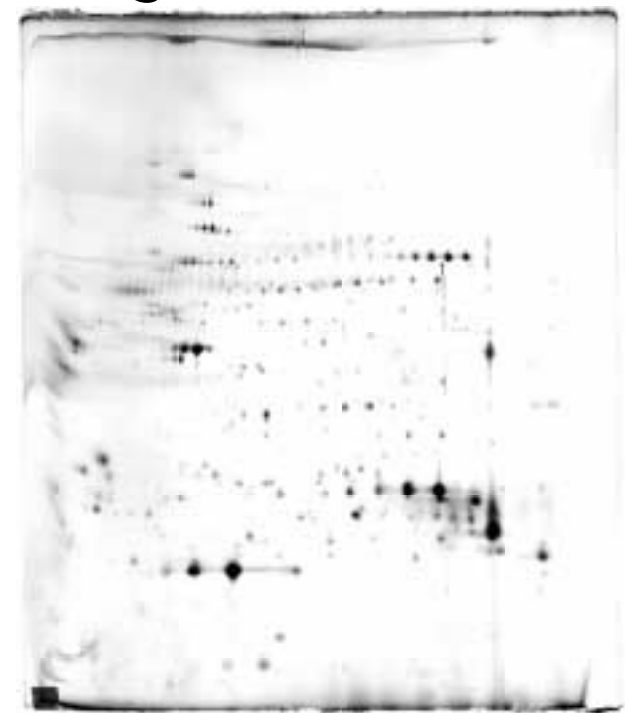

Figure 1

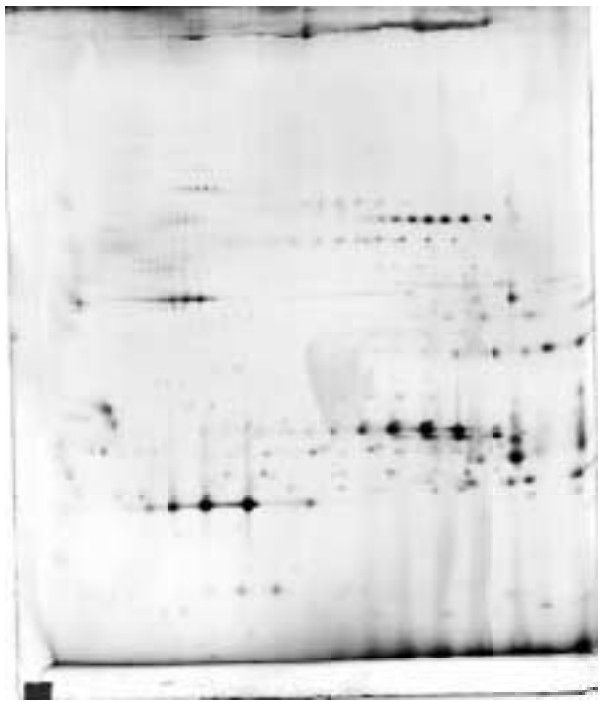 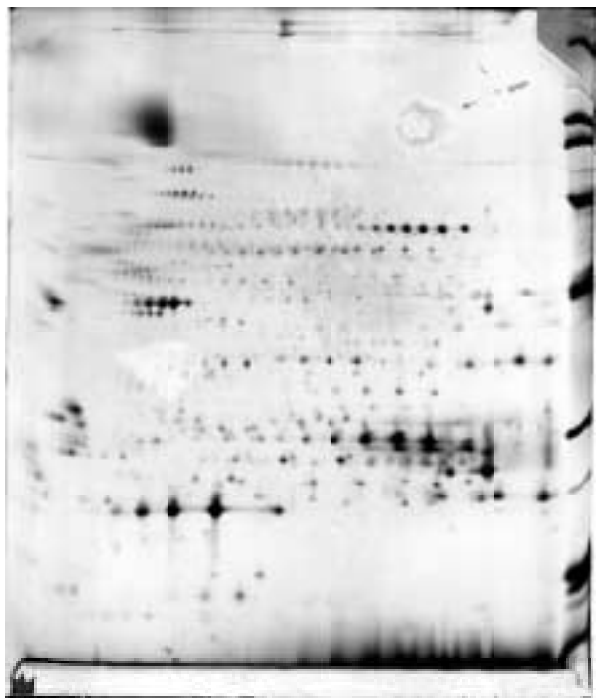 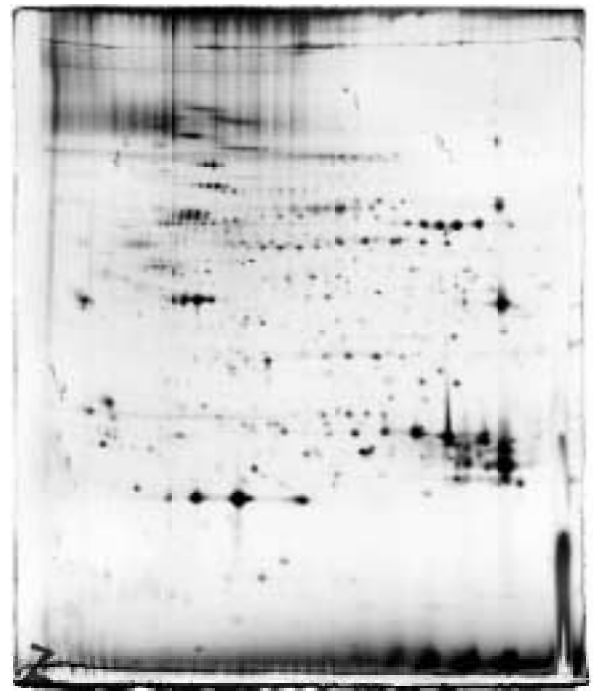 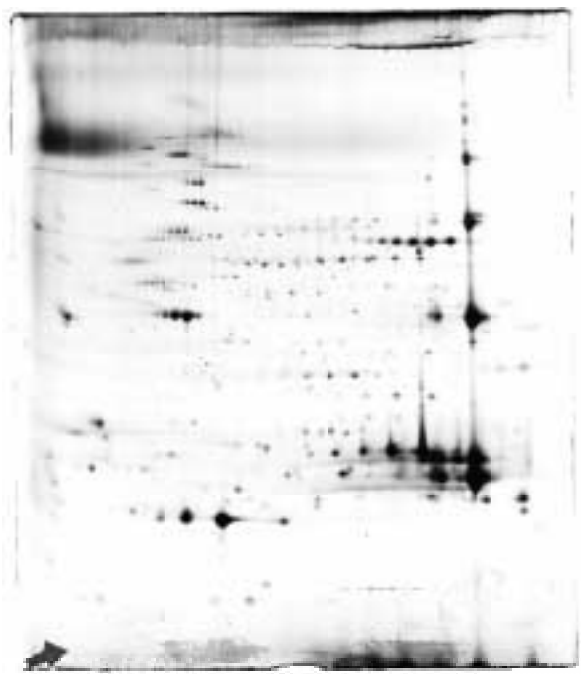 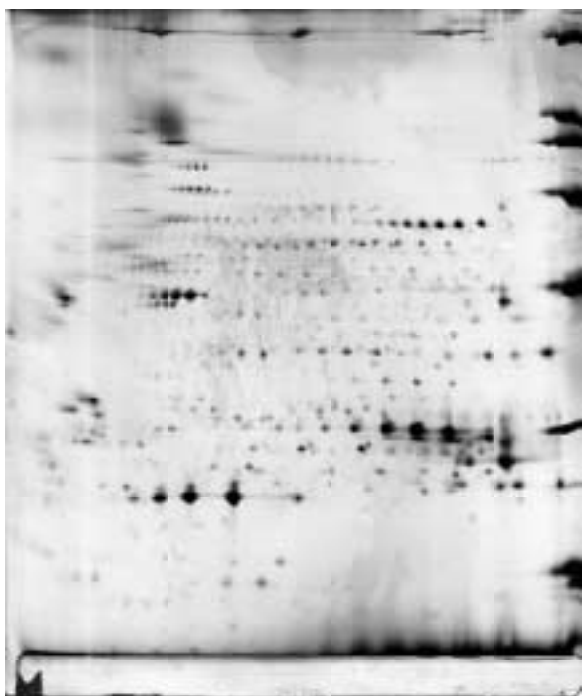 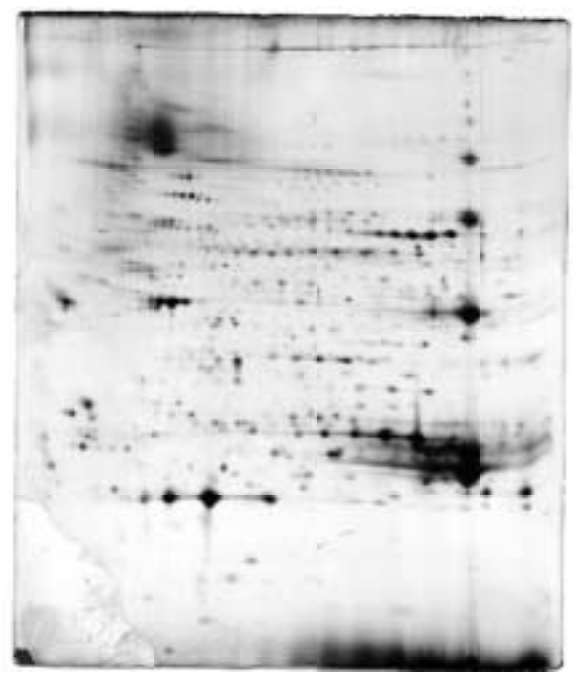

**Figure 2**

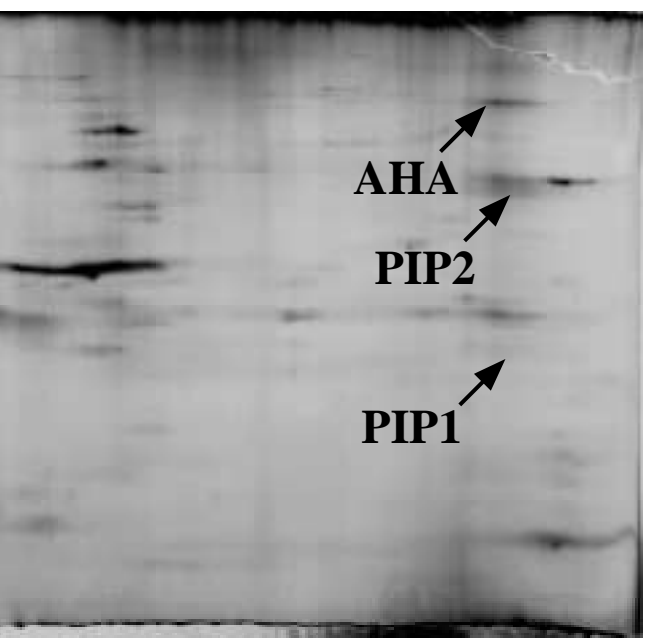
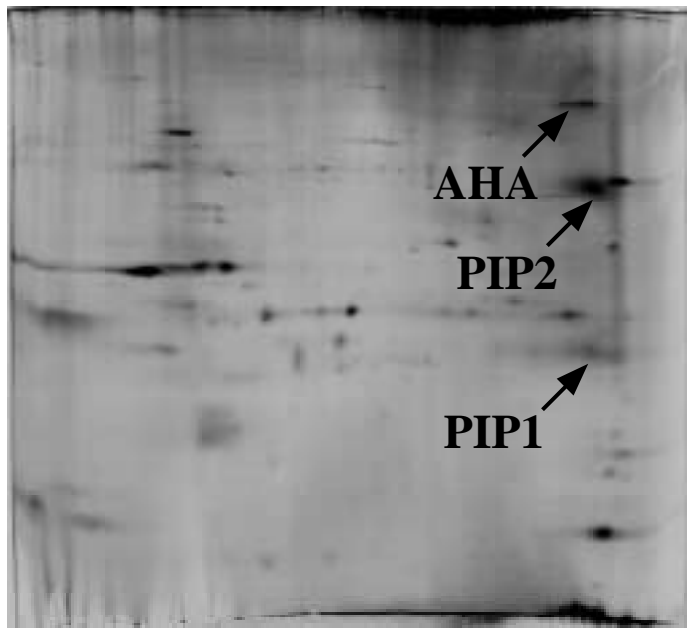
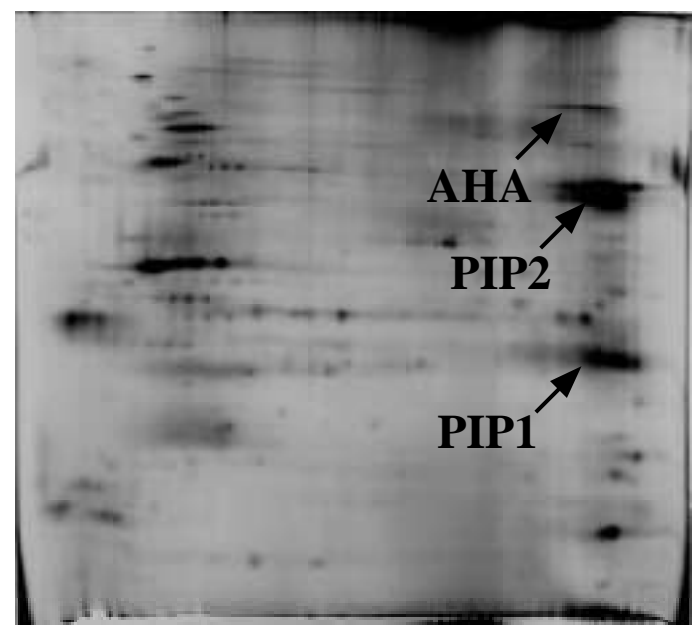
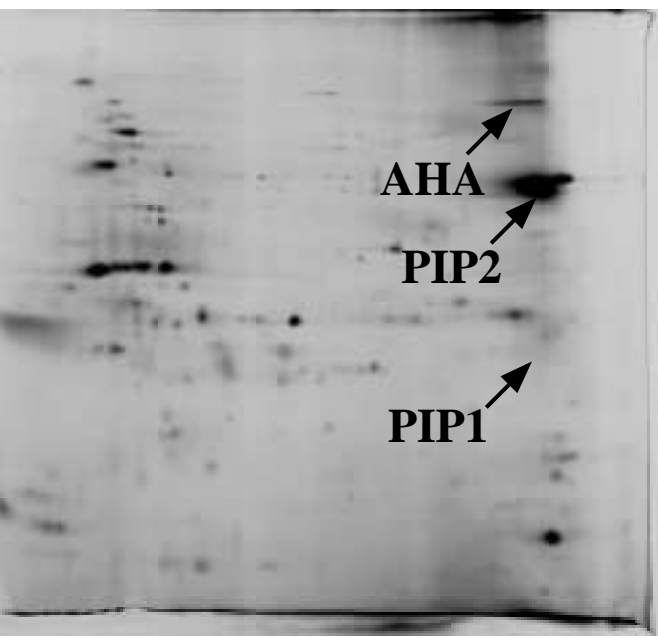
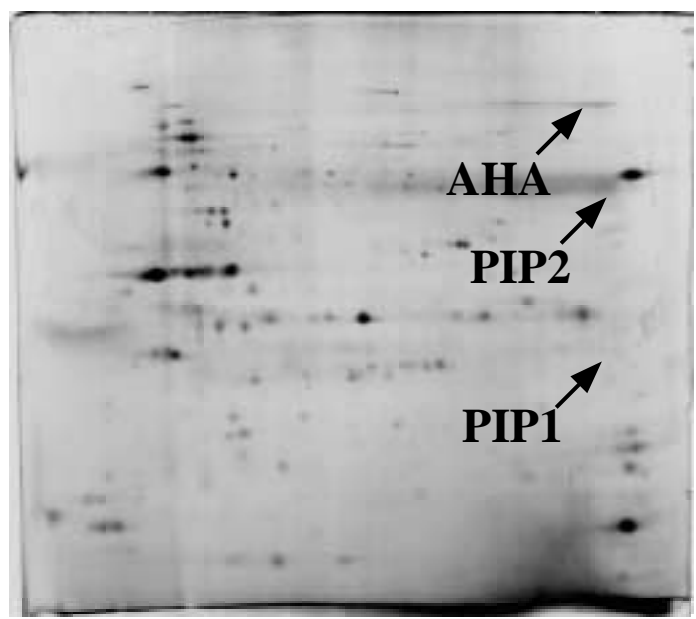
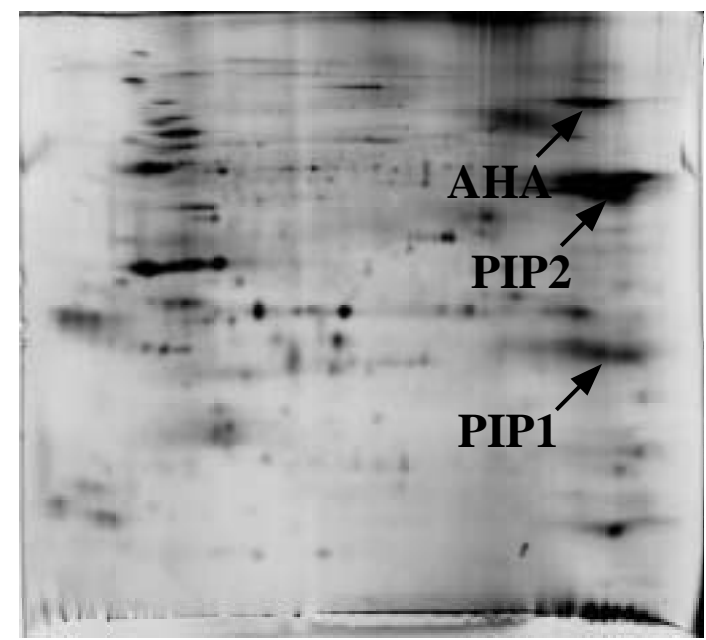

**Figure 3**